**Rupture index: a quantitative measurement of sub-micrometer cracks in graphene**

*Hadi Arjmandi-Tash, Lin Jiang and Grégory F. Schneider\**

Dr. H. Arjmandi-Tash, L. Jiang and Dr. G. F. Schneider
Leiden University, Faculty of Science, Leiden Institute of Chemistry, Einsteinweg 55,
2333CC Leiden, The Netherlands

E-mail: g.f.schneider@chem.leidenuniv.nl

Per weight graphene is stronger than steel, but because graphene is so thin, it breaks and evidently forms cracks during handling, particularly when supporting polymers are not used for transfer. In this paper, the parameter 'rupture index' is introduced to quantify the density of the cracks and related defects in graphene during processing. The approach takes the advantages of the high contrast achievable in fluorescence quenching microscopy to distinguish between graphene and the background substrate, visible through the cracks in graphene. The rupture index can well compare the effectiveness of different graphene transferring methods in minimizing the formation of the cracks, and quantify the oxidation of metals protected with graphene.

## 1. Introduction

The abrupt interruption of the honeycomb structure at the edges and at crystalline defects alters the local electrical and chemical properties of graphene [1]. Several techniques including scanning tunneling microscopy [2–4], transmission electron microscopy [5,6], electron energy loss spectroscopy [6,7] and Raman spectroscopy [8] have been used to characterize edges and defects in graphene. Those techniques are local and can hardly provide any global and quantitative measure of the density and topology of the edges in large scale, which is of great importance in practice.

Graphene quenches the emission of fluorophore molecules placed in close enough vicinities: available electronic states in graphene promote non-radiative relaxation (no photon emission) of electrons at excited states via a long-range resonance energy exchange mechanism [9,10]. The principle, called *fluorescence quenching microscopy* (FQM), allows the visualization of the *shadow* of graphene [11,12]. In optimized conditions, the intensity of the quenching (the darkness of the shadow) depends on the number of layers of graphene, hence the technique is capable to make distinctions particularly between monolayer and bilayer graphene [12]. FQM is compatible with various fluorophores and most nanofabrication techniques. Particularly, hydrophobic fluorophores can be mixed with poly(methyl methacrylate) (PMMA), as the conventional polymer used for transferring or patterning graphene. All such capabilities turned FQM as a general visualization method for quick observation and characterization of graphene and related materials [13,14].

In this article we define the parameter "rupture index (RI)" to quantify the average density of the cracks in graphene based on fluorescence quenching microscopy, and demonstrate that cracks with a separation between the edges as small as ~ 650 nm are versatile to detect. By solubilizing fluorescent molecules in a polymeric matrix, which is then spin-coated on the surface of graphene, the rupture index can be measured. A series of graphene samples transferred using different methods yield different RIs. A RI value can be determined for any type of graphene and particularly in situations where graphene is used as a protective coating, for example in preventing the oxidation of metallic surfaces.

Determination of the RI is independent of the experimental conditions, an enormous asset for the comparison of the quality of all the graphene types now manipulated worldwide. FQM is sensitive to the type of cracks generated in graphene, particularly for differentiating between cracks induced mechanically or chemically.

## 2. Results

### 2.1. Crack visualization in the graphene lattice

**Figure 1** shows the same location of a graphene sample observed using an optical microscope (Figure 1-a), a scanning electron microscope (Figure 1-b), and a fluorescent microscope (Figure 1-c). The visibility of graphene is the lowest in optical micrographs and improves slightly in the electron micrograph. Fluorescent quenching microscopy, however, provides the best visibility and highest contrast between graphene and the supporting substrate. The grayscale intensity histograms of the images are plotted below the corresponding image. The histograms for the optical and electron micrographs are very sharp (sharper for the optical micrograph) and localized, which justifies the low contrast. The pixels of the fluorescent micrograph image, however, are split to either very dark (on graphene) or very bright (on cracks) extremes, which allows distinguishing unambiguously the cracked regions from pristine graphene.

The stability in time of the fluorescence quenching microscopy (FQM) images over scanning electron microscopy (SEM) images represents another important advantage of FQM. In fact, during SEM imaging, the surface of the sample exposed to the electron beam changes in color as a result of the interaction of the material with the electron beam. Such an exposed area with modified color (darker) is shown in the top right corner of the SEM image in Figure **1** 1-b. We note that optical microscopy provides the best visibility of graphene only on silicon oxide layer of ~285nm [13]. Application of electron microscopy also is limited to the samples in which graphene is placed on electrically conducting materials; the efficiency of FQM, however, is independent of the substrate properties.

In order to further monitor the sensitivity of FQM towards imaging of cracks, we next deliberately crumpled a copper foil containing graphene and compared the results with an aged graphene sample (~6 months old) kept in air at room temperature (**Figure 2**Figure 2-a). FQM shows that the defects generated during the mechanical crumpling are visible in the form of straight lines (Figure 2-b). The majority of the defects in the aged graphene sample, are of rounded shapes (Figure 2-c), suggesting that such defects may have nucleated at central points and grown over the time isotropically.

## 2.2. Resolution of FQM for crack visualization

Figure 3-a shows the FQM of microfabricated graphene ribbons of respectively 1μm, 2μm and 3μm in widths (an SEM image of the widest graphene ribbons is shown in Figure 3-b). In order to determine the resolution of FQM in characterizing cracks in graphene, we plotted the grayscale intensity along the dashed line in Figure 3a and through the ribbons (Figure 3-c). Oscillations of the grayscale intensity correspond to the emission of fluorophores, periodically turned 'on' and 'off'. Near the edges of the graphene ribbons, the intensity of the gray channel changes at a rate independent of the width of the ribbons. By overlaying the intensity oscillations for the three sets of ribbons (Figure 3-d), and fitting the slope of tangent at the inflection point, we determined the lateral resolution of the method to be around 650nm.

## 2.3. Rupture index

The sharp contrast between on/off graphene areas in FQM images provides the possibility to numerically quantify the cracks in graphene. For that, we transferred graphene sheets using three different approaches where i) a polymeric resist (PMMA) [15], ii) a lateral frame and iii) no support were used to transfer graphene (see schematics in **Figure 4**). The presence of the PMMA resist prevented cracking of graphene during etching the copper foil, as shown by the fully quenched fluorescence in the FQM micrograph (Figure 4-a). For the two other transfers, however, cracks in graphene are largely visible as white areas (Figure 4-b and c).

We developed a MATLAB script (see supporting information) to process FQM micrographs and to identify the pixels corresponding to the borders of the cracks. **Table 1**Table 1 details the process within the script. Bottom panels in

Figure 4-a, 4b and c shows the FQM images processed with our script. The white pixels, now, correspond to the borders of the cracks in graphene. Using the processed images, we define the '*rupture index (RI$_p$)*' as the total number of the white pixels (lying at the perimeter *p* of the cracks) in the image divided by the total number of pixels corresponding to graphene, multiplied by one thousand.

The rupture index RI$_p$ provides a measure for the density of the ruptured graphene areas. We calculated the RI$_p$ for eight different windows in each sample; the results are plotted in Figure 4-d. PMMA assisted, frame stabilized and bare graphene transferring methods yielded average RI$_p$ of 1.4, 27.1 and 41.5 respectively.

We note that FQM images are of negligible sensitivity to small changes in the resist thickness, and to the presence of wrinkles, folds and overlaps. For example, if the thickness of the fluorescent resist is less than ~630nm (Figure S3), the visibility of the cracks are not much affected by the variations of the thickness of resist. Similarly, the wrinkles visible in the SEM image in Figure 3-b do not affect the contrast in the corresponding FQM image. The simplicity of the technique (that is spin coating a resist and imaging it using a fluorescence microscope), the fast processing (to be compared with techniques such as atomic force microscopy), the possibility for automation for large amount of data and the accessibility of the materials (PMMA, fluorophores) are also remarkable. Accordingly, rupture index represents a new scale to characterize the density of the cracks in graphene.

## 2.4. Quantifying the efficiency of a graphene coating to prevent oxidation of metals

Graphene possess remarkable advantages to prevent the oxidation of metals [16]: i) mono and multilayers of graphene can be grown straightforwardly on a wide range of metals [17]; ii) full coverage can be achieved on the metallic parts with complicated geometries (e.g. very small holes) which are not easily accessible for other oxidation prevention methods [18]; iii) unlike the methods such as painting, graphene is atomically thin and therefore preserves the 3D topology of the coated parts [19]; and iv) the chemical and thermal stability of graphene provides efficient oxidation resistivity under different chemical environments and in a wide range of temperature [20][21].

Rupture index can quantify the efficiency of graphene-based coating in preventing metal oxidation. For this, we prepared three graphene samples on copper foils. Two of those samples were increasingly crumpled (**Figure 5**-a) to deliberately generate cracked graphene coatings with two different crack densities. A forth sample, composed of a bare copper foil (without a graphene coating) was used as a control representing the maximum oxidation level. A part of the graphene for each sample was transferred onto a silicon wafer to measure the corresponding rupture index. As the oxidation depends on the surface $s$ of the cracks, we refine the rupture index (now called surface rupture index, $RI_s$) as the total number of the pixels located at the surface of the cracks divided by the total number of pixels forming the image, again multiplied by one thousand. We estimated the surface rupture index to be $RI_s$= 3.2, 127.7 and 220.8 for the three samples respectively. With this definition, the bare copper foil has $RI_s$=1000.

Next, $RI_s$ values were associated to the redox behavior of ferricyanide measured by cyclic voltammetry (CV) curves. For CV measurements, we used the samples directly as the working electrodes in an electrochemical cell filled with an electrolyte solution composed of $K_2CO_3$ (100 mM) and of the redox probe ferricyanide (5 mM) (Figure 5-b). In such graphs, the current peaks (the distance from the maximum of the peak to the tangent baseline), at 0.6 V and in the range of -0.2 V ~ -0.4 V are attributed to the oxidation and reduction of ferricyanide, respectively. During the experiments, uncovered areas of copper (through the cracks in the graphene coating) began to oxidize into $Cu^+$ and further reacted with the oxygen reduction product OH⁻ (around 0.4 V ~ 0.5 V) to form electrochemical-inert $Cu_2O$ on the surface of Cu foil. For increasing $RI_s$ values, the continuous replacement of Cu with the red $Cu_2O$ during the CV measurements yields: i) a decreasing electron transfer rate of the electrode which gradually lowers the redox peak current values, and ii) an increasing intensity of the redish color of the copper underneath graphene (Figure 5-a, bottom panel). Additionally, the current drop between the initial and final state for same experiment duration is higher for samples with cracked graphene (see Figure S4).

To quantify the loss of the copper electrochemical activity, we define the current degrading rate (CDR) as depicted in **Equation (1)**:

$$CDR = \frac{I_{initial}(\mu A) - I_{final}(\mu A)}{experiment\ duration\ (s)} \qquad (1)$$

The CDR value can be related to the amount of red copper oxide forming as CV measurements undergo. Therefore, a higher CDR is expected for samples containing less uniform coatings (i.e., more cracks). After ~30 minutes of continuous CV measurements, the dependency of CDR vs RI is best fitted using an exponential equation (Figure 5-c):

$$CDR(RI_s) = 1.69 - 1.59e^{-\frac{RI_s}{512,8}} \qquad (2)$$

CDR achieved in this equation is in unit of μA/s. Accordingly, the normalized anti-oxidation resistivity ($\mathcal{R}$) of graphene coatings with different surface rupture indexes can be calculated as:

$$\mathcal{R} = 1 - \frac{CDR(RI_s)}{CDR(1000)} \qquad (3)$$

$\mathcal{R}$ is a unitless parameter and ranges between 0 (for bare copper foil, $RI_s$=0) to 1 (for samples with leak-tight coating, $RI_s$=1), Figure 5-c. Once cracks are generated in graphene, $\mathcal{R}$ decreases exponentially if plotted versus $RI_s$. Note that equation 1 predicts a finite level of oxidation even at $RI_s$ equal to zero; indeed crystalline defects and cracks smaller that the resolution of the technique (less than 650nm in width, see above) can still account for this minimum oxidation in CVD graphene. Interestingly, a graphene piece with $RI_s$=512.8 (almost 50% graphene coverage) shows only 24% of the anti-oxidation resistivity compared to a crack free graphene.

## 3. Conclusion

Measuring the influence of the macroscopic edges on the global properties of graphene has always pended on the development of a powerful tool to quantity cracks. We introduced a quantity − the rupture index − which serves well to estimate the amount of the microscopically visible openings in graphene. The parameter is robust to small changes in the measurement conditions and can be used to quantify the presence and amounts of cracks in graphene universally. The measurement technique benefits from the ultra-high contrast between on/off graphene areas achievable in the fluorescence quenching microscopy and is fully automatable. We show that rupture index offers a solution for quantitatively analyze and predict the chemical reactions associated to the formation of cracks in graphene such as the oxidation of the underlying copper foil.

## 4. Experimental Section

We use the graphene chemically grown on copper foil (Alfa Aesar, 99.999% purity, 25µm thickness) in a cold wall CVD set-up. Graphene covers the surface of the copper continuously. For fluorescent quenching microscopy experiments, we transfer this graphene onto a silicon wafer with ~285nm thermally oxidized capping layer. We used well-established recipes for transferring graphene [15], unless we pre-mixed $2\ \mu l - 6\ \mu l$ of $4\ mM$ solution of Rhodamine B (dissolved in acetone) with $100 \mu l$ of the supportive poly(methyl methacrylate) PMMA layer and used during transferring.

FQM was performed on a Axiovert 200 ZEISS inverted fluorescence microscope equipped with a monochrome AxioCam MRm ZEISS camera. The camera is both manually and automatically adjustable, but we always used manual adjustments and fixed all the parameters the same for all the images. We note that the quenching is observable even in the eyepiece of the microscope and thus the principle is independent the setting of the camera.

**Supporting Information**

Supporting Information is available from the Wiley Online Library or from the author.

**Acknowledgements**

The work leading to this review has gratefully received funding from the European Research Council under the European Union's Seventh Framework Programme (FP/2007-2013)/ERC Grant Agreement n. 335879 project acronym 'Biographene', and the Netherlands Organisation for Scientific Research (Vidi 723.013.007). The authors are grateful to L. Macedo Coelho Lima for the required trainings on the FQM setup and for scientific discussions.

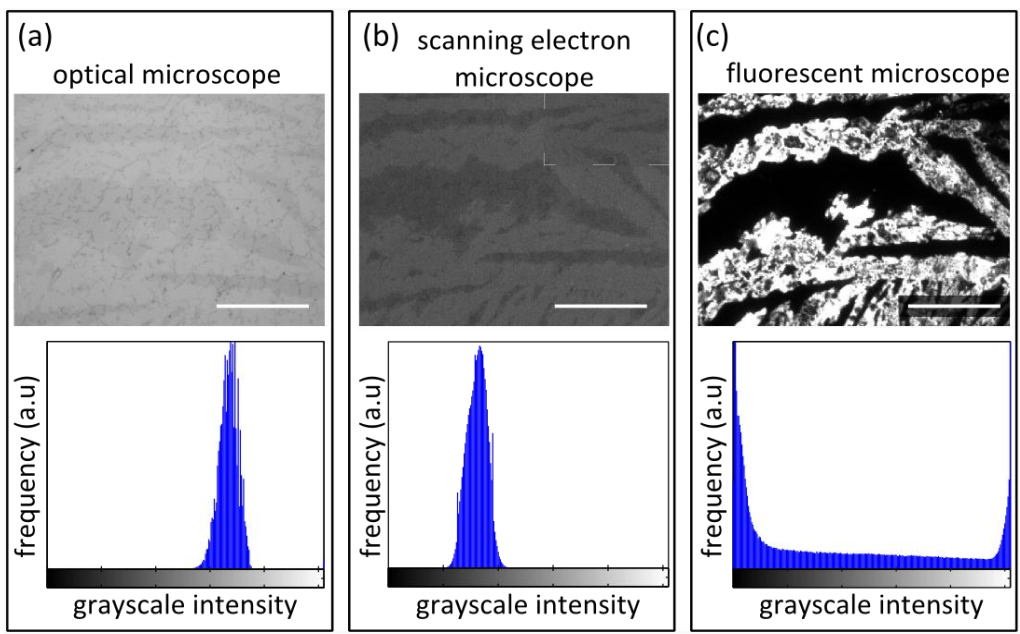

**Figure 1: Efficiency of fluorescence quenching microscopy (FQM) for the visualization of the cracks in graphene**

**a)** Optical image (top panel) of a cracked area in graphene transferred on a silicon wafer with ~285 nm of a silicon oxide capping layer; bottom panel shows the histogram of the grayscale intensity of the image.

**b)** Scanning electron microscope image of the same area as in (**a**); the image is captured with an electron beam of 5 kV. A rectangular area with darker color (previously exposed by electron beam) is detectible in the top-right corner of the image. Bottom panel shows the histogram of the grayscale intensity of the image.

**c)** Fluorescent microscope image of the same area as in (**a**) and (**b**); graphene was covered with PMMA mixed with Rhodamin B solution (100:2 ratio). The image was taken with an exposure time of 2ms. Bottom panel shows the histogram of the grayscale intensity of the image.

All the scale bars correspond to 50 μm.

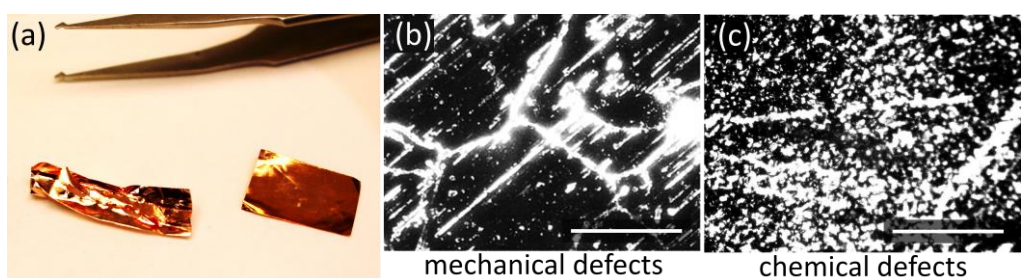

**Figure 2: Comparison of the FQM of mechanically and chemically generated defects in graphene**

**a)** Photograph of two graphene on copper samples. Left: freshly grown graphene that was intentionally damaged by crumpling the copper foil. Right: a graphene sample stored ~6 months at room temperature in air; due to the oxidation, the copper foil looks more reddish than normal.

**b)** FQM micrograph of the mechanically defected graphene (left sample in **a**) after transfer on an oxidized silicon wafer.

**c)** FQM micrograph of the aged graphene (right sample in **a**) after transfer on an oxidized silicon wafer.

All the scale bars correspond to 50 μm.

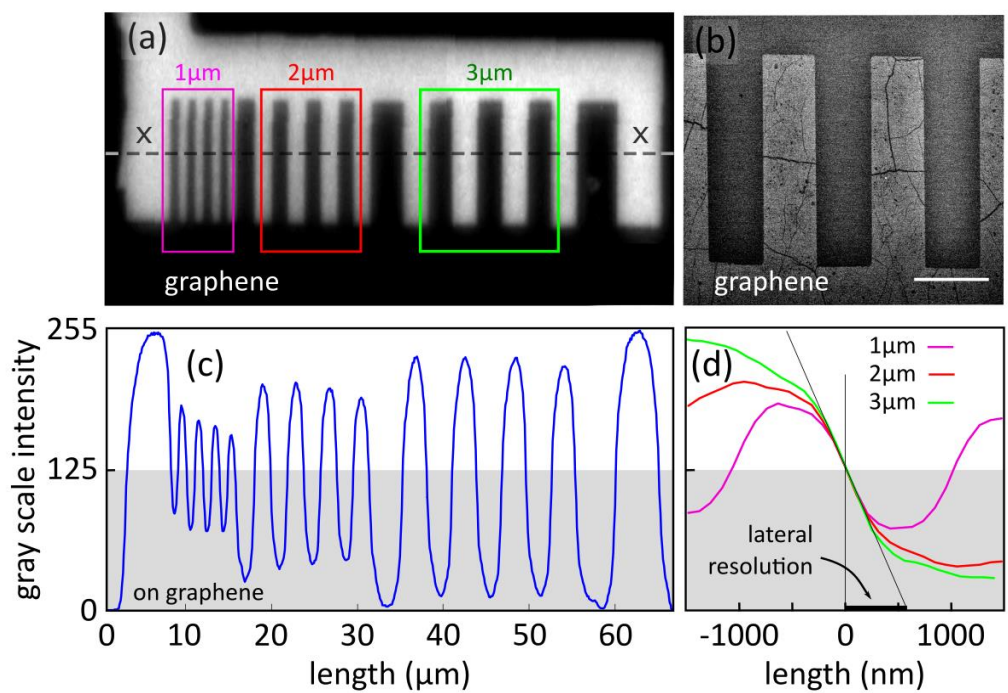

**Figure 3: The lateral resolution of FQM**

**a)** FQM of graphene ribbons with 1μm, 2μm and 3μm widths: For each set, the gap between the ribbons are the same as their widths.

**b)** Scanning electron microscopy image of the widest ribbons (green window in **a**). The scale bar corresponds to 4 μm.

**c)** Grayscale intensity along the line x-x in **a**

**d)** Overlaid grayscale intensity oscillations corresponding to ribbons of different widths in **c**. The vertical axis has the same unit as in **c**.

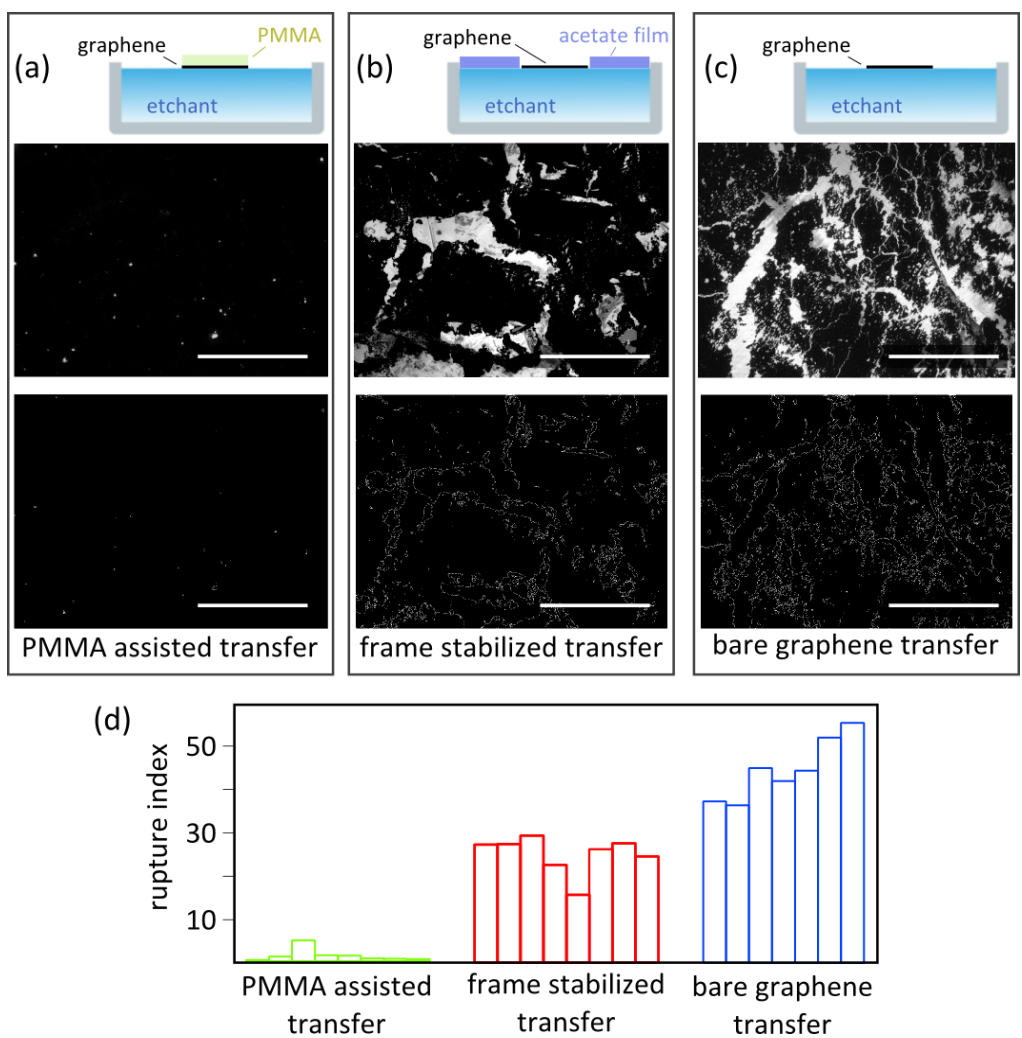

**Figure 4: Fluorescent microscopy for the visualization of cracks in graphene**

**a)** Schematic of the method (top panel) and FQM image of a selected area (middle panel) and processed image (see the text, bottom panel) corresponding to graphene transferred with a PMMA support

**b)** The same as *a* for a graphene piece transferred with lateral stabilization during etching and fishing

**c)** The same as *a* and *b* for a graphene piece transferred without any support

**d)** Rupture index calculated for eight different windows inside the samples discussed in a-c.

All the scale bars correspond to 50 μm.

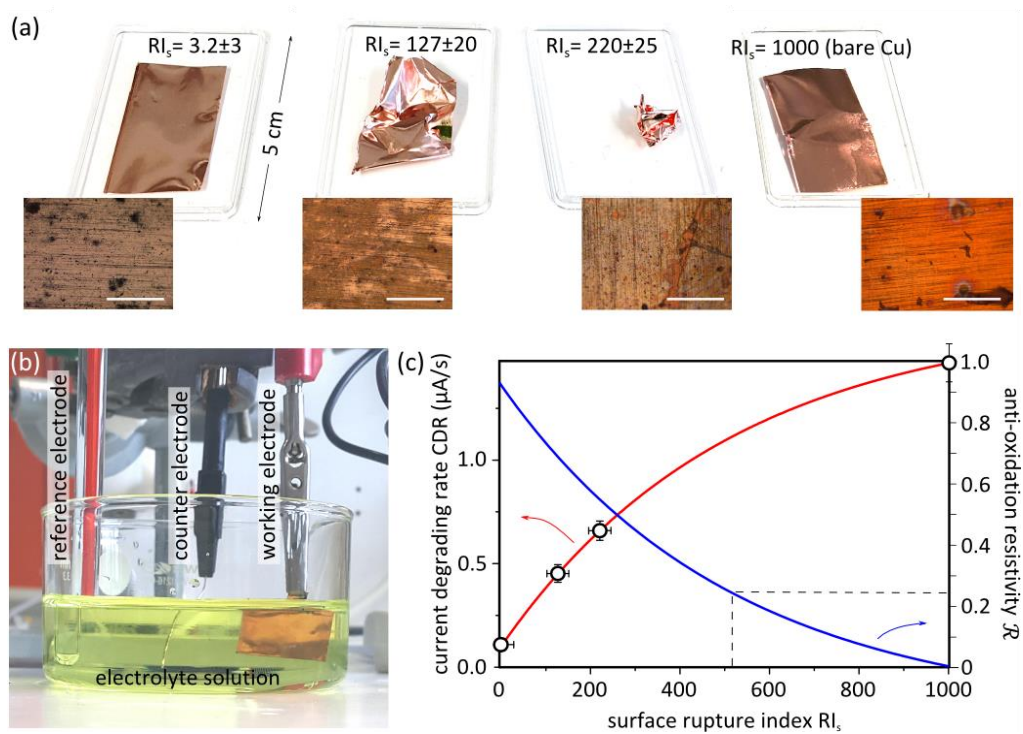

**Figure 5: Rupture index to study the oxidation level of copper foils coated with graphene**

**a)** Photographs of the experimental samples: Copper sheets coated with graphene samples with different surface rupture indexes (increasing from left to right) were prepared by mechanically crumpling copper foils. The sample in most right is bare copper foil. Bottom optical micrographs represent the oxidized samples at the end of the experiment. The scale bars show 400 µm.

**b)** Photograph of the experimental setup showing the three-electrode configuration with Red Rod reference system and Pt as the counter electrode. The electrolyte solution is 100 mM K2CO3 with ferricyanide (5 mM) as the redox probe.

**c)** Current degrading rate (CDR) and anti-oxidation resistivity ($\mathcal{R}$) for the copper samples presented in **a.**

**Table 1:** Standard process for the calculation of the rupture index

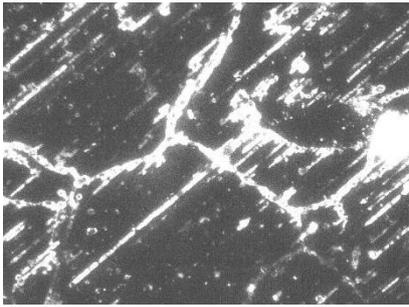

**Step 1: Spin coating the sample, capturing the image**

Spin-coat a suitable amount of PMMA (AR-P 662.06 ALLREIST) in which 4mM of Rhodamine B solution (in acetone) is dissolved (2:100 ratio).
The FQM images are captured with 20X objective in the normal operation condition of a fluorescent microscope. Even-though the FQM images are of high contrast, fine-tuning the contrast of the images at this step may further help to improve the accuracy.

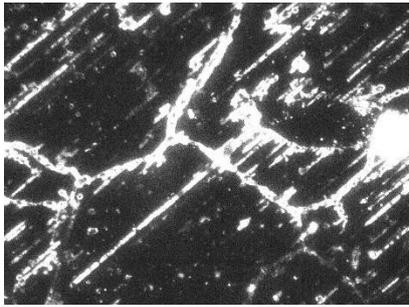

**Step 2: Normalization of the image (optional)**

The goal of this step is to eliminate the effect of any intensity variations from image to image. The intensity of each pixel of the image after this transformation ($y$) is calculated by:

$$y = 255 \times \frac{(x-min)}{(max-min)}, \text{ where:}$$

$x$: the gray scale initial intensity of the pixel
$min$: minimum gray scale intensity for the original image.
$max$: maximum gray scale intensity for the original image.

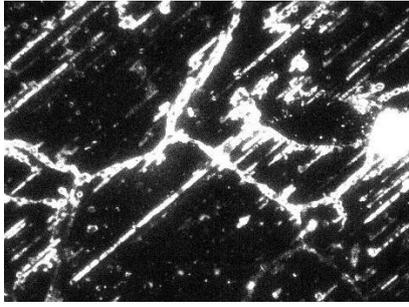

**Step 3: Adjusting of the contrast**

The goal of this step is to further improve the contrast of the image and increase the visibility of small cracks.
Built-in $Imadjust()$ function of the MATLAB performs this task by saturating 1% of data with the lowest and highest intensities.

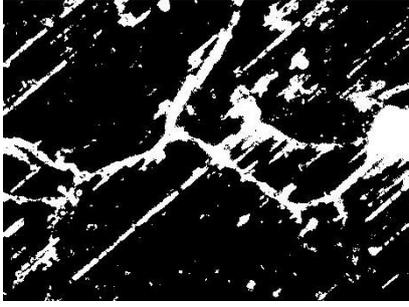

**Step 4: Conversion the *gray-scale* into the *binary* image**

With this conversion, all the pixels with the intensity greater than the median (i.e. 255/2) will be replaced by 1 (white color, corresponding to the cracks) and the pixels with the intensity less than the median will be replaced by 0 (black, corresponding to graphene).

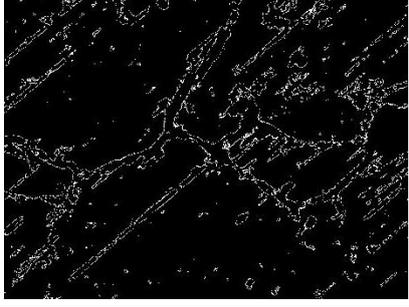

**Step 5: Finding the borders of the cracks, calculating rupture index**

By using the built-in $bwperim()$ function of the MATLAB, we can find the pixels corresponding to the borders of the cracks. This function returns the pixel which are white, but are connected to 4 black neighboring pixels.
The rupture index of the image at this step is calculated as the total number of the white pixels (borders of the cracks) normalized by the number of the black pixels in the image of step 4.

# Supporting Information

**Rupture index: a quantitative measurement of sub-micrometer cracks in graphene**

*Hadi Arjmandi-Tash, Lin Jiang and Grégory F. Schneider\**

**Supporting information**

**Principle of fluorescence quenching by graphene**

Some molecules exhibit fluorescence properties: Here, an electron on a ground state is excited to a higher energy level upon the absorption of the energy of an impinging photon (shown as the green arrow in the left panel of the **Figure S1**-a); the electron at high energy levels may loss its energy via few non-radiative relaxation events (blue arrow). In a fluorescent molecule, however, a remarkable relaxation is via radiation of a photon (yellow arrow). In the presence of the conductive material (e.g graphene, right panel in a), the excited electron can offer its energy to the available electrons in the material; hence relaxation is without any emission. The principle is called quenching.

In our experiments, we dissolved fluorescent Rhodamine B molecules in PMMA layer (Figure S1-a). The efficiency of dye-graphene energy exchange is distance dependent: the emission of the dye molecules close to graphene are well quenched. Far from the graphene, however, quenching will be less efficient.

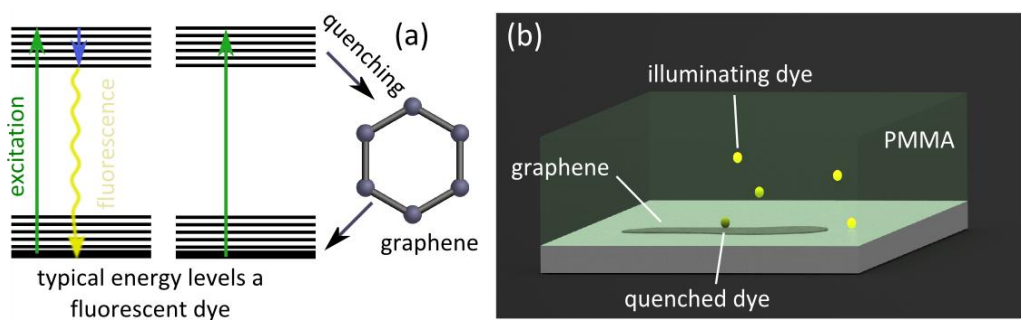

Figure S1: Principal of fluorescence quenching

a) Electronic-state diagram showing the process of resonance energy exchange during a fluorescence emission (left) and quenching (right) process
b) Fluorescence quenching as a function of the distance to a quencher (graphene): dye molecules are immersed in a polymeric matrix (PMMA in this case). While the dye molecules far from graphene illuminate normally, the illumination of the molecules close to graphene are quenched.

**The role of the polymeric matrix**

We found that the presence of the polymeric matrix (PMMA in our experiment) is very important to achieve uniform coverage of the dye molecules. F**igure S2** shows examples of the non-uniform coating achieved by direct spinning Rhodamine B dissolved in organic solvents.

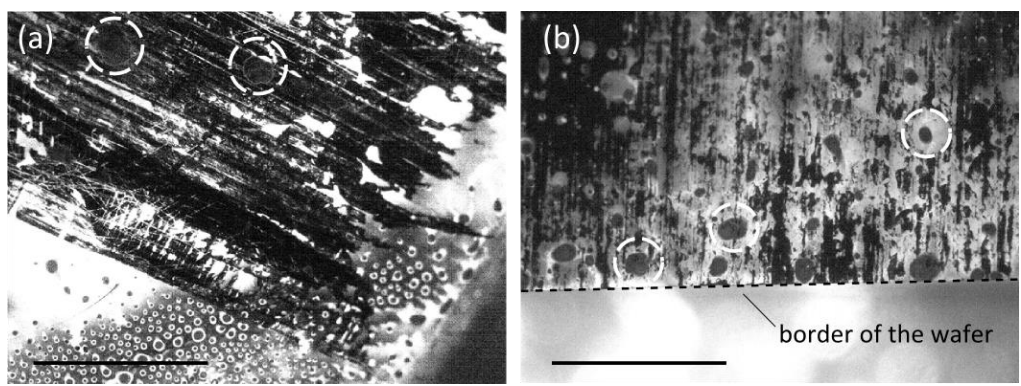

Figure S2: Non-uniform aggregation of the fluorescent dye without using the polymeric matrix

Images were taken upon direct spin-coating 250μM (a) and 130μM (b) of Rhodamin B in acetone solution on the samples. Few of the spots showing non-uniform aggregation of the dye are marked on the figures. Scale bars correspond to 50μm.

**Effective thickness of the dye-PMMA coating**

**Figure S3** shows the experimentally measured effect of the thickness of the polymeric matrix. We first spin-coated the sample by the dye-PMMA layer of 300nm. Graphene highly quenches the nearby dye molecules, hence looks darker than the surrounding. The contrast between on/off graphene areas lowered when we increased the thickness of the layer to 630nm (by spin-coating another layer on top). The brighter color for graphene is due to the emission of the dye molecules placed at the higher levels of the coating. By increasing the thickness to ~1 $\mu m$ the graphene became fade to a large extent.

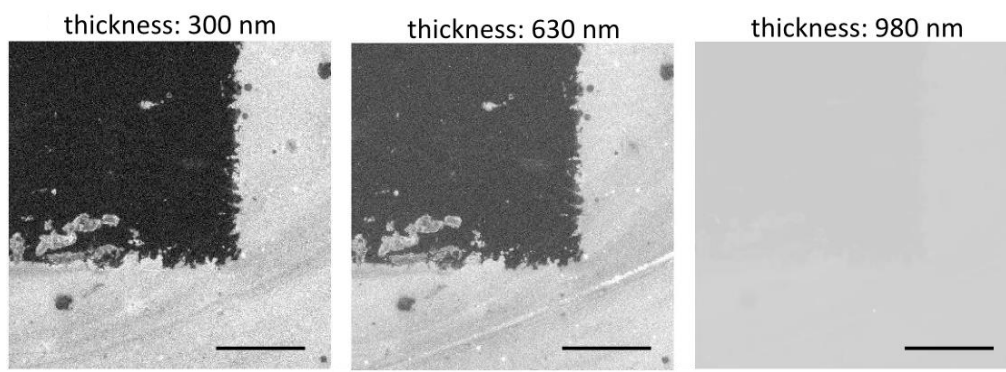

Figure S3: Fluorescence quenching microscopy of a same window of a graphene/SiO2 sample covered with different thicknesses of the polymeric matrix; The scale bars correspond to 200μm.

**MATLAB script to calculate the Rupture Index**

```
% This is a MTLAB script to calculate the Rupture index
of the FQM images.

%% call files and convert to tiff

n=12; %total number of the FQM images
index=zeros(n,1);

for m=1:n

filename=['L25_20X_0' num2str(m)]; %adjusting the image
name

disp(filename);
image=imread(filename, 'jpg'); %reading the images
image = imadjust(image); %improving the contrast

%% converting the gray figure into a BW figure

coeff = 1;  %adjustment coefficient
level = graythresh(image)/coeff;
BW = im2bw(image, level);

%% Calculating perimeter of the cracks
BW2 = bwperim(BW);
imwrite(BW2, [filename,'_prm.tif']);
```

```
[a,b]=size(BW2);
perimeter=0;
black =0;

for i=1:a
    for j=1:b
        if (BW2(i,j)==1)
        perimeter=perimeter+1;
        end
        if (~BW(i,j))
            black=black+1;
        end
    end
end

%% saving data
index(m)=1000*perimeter/black;
end
bar(index,'DisplayName','index');
```

**Electrochemistry as a measure of oxidation resistance**

Electrochemical experiment provides quantitative data about the oxidation status of the samples. Cyclic voltammetry (CV) curves reporting the electrochemical behavior of the samples at the initial state when just immersed in the ferricyanide solution (**Figure S4**-a) and after ~30 minutes continuous CV tests (Figure S4-b). Potassium ferricyanide was used as the redox probe here to monitor the electrochemical performance of graphene coated Cu as working electrode.

In the electrochemical measurement, Cu acts as the "active" working electrodes with different coverages of graphene. According to the standard redox potential, the oxidation of copper into Cu+ overlaps with the oxidation peak of ferricyanide. And then react with hydroxyl ions that are reduced from oxygen dissolved in the solution to generate red copper oxide:

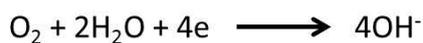

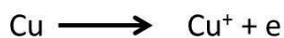

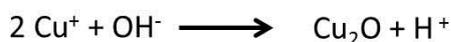

The reduction peak in the range of 0.4 – 0.5 V (Figure S4, both a and b) that can be ascribed to oxygen reduction and the observed red color on Cu foil both confirmed this oxidation process. With the oxidation process continued, the coverage of copper (I) oxide on Cu foil increased gradually and partially replaced the original copper as the working electrode. As a result, the current signals began to degrade owing to the poorer electron-transfer ability of copper (I) oxide compared to copper. Figure S4-a and b corresponding to the initial and final cyclic voltammetry measurements for Cu with different coverage of graphene confirmed the prediction. The CV current signals of final tests, especially, manifest the effectiveness changes of graphene protection with the variable coverages. In details, compared with the quasi-reversible redox peaks of black curve (RIs as 3.2), the red and blue curves with higher RIs values (127.7 and 220.8, respectively) exhibited much weaker signal of the redox peaks. In short, the electrochemical results affirm that graphene coverage on Cu foil protects copper from oxidation, and the better coverage (smaller RIs values) gives better protection.

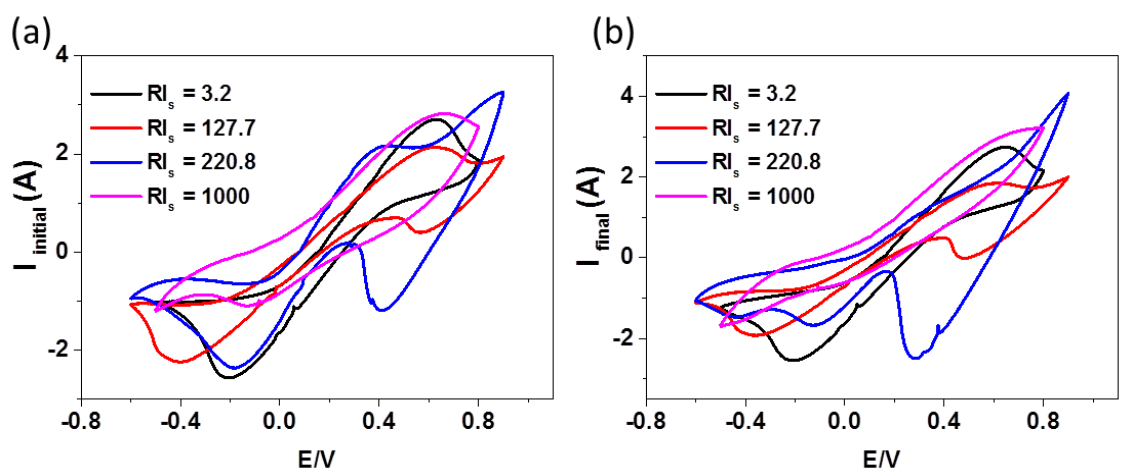

Figure S4: Initial (a) and final (b) current-potential (I-V) curves acquired from graphene samples with different rupture indexes in ferrocyanide/ferricyanide solution (80 CV cycles at a scan rate of 100 mV/s were performed between the initial and final conditions).